\documentclass[letterpaper]{article}
\usepackage[top=3cm,bottom=3cm,left=3cm,right=3cm,marginparwidth=1.75cm]{geometry}
\usepackage{listings}
\usepackage{xcolor}
\usepackage{cite}
\usepackage{multirow}
\usepackage{amsmath}
\usepackage{amssymb}
\usepackage{dcolumn}
\usepackage[english]{babel}
\usepackage{xspace}
\usepackage[utf8x]{inputenc}
\usepackage[T1]{fontenc}
\usepackage{palatino}
\usepackage{dsfont}

\usepackage{amsmath}
\usepackage{afterpage}
\usepackage{graphicx, subcaption}
\usepackage[colorinlistoftodos]{todonotes}
\usepackage[colorlinks=true, linkcolor=blue, citecolor=blue]{hyperref}

\newcommand{\be}{\begin{equation}}  
\newcommand{\ee}{\end{equation}}  
\newcommand{\bea}{\begin{eqnarray}}  
\newcommand{\eea}{\end{eqnarray}}  
\newcommand{\ud}{\mathrm{d}}
\newcommand{\cmtwo}{cm$^2$\xspace}

\newcommand{\vv}{\mathbf{v}}
\newcommand{\uu}{\mathbf{u}}

\newcommand{\rr}{\mathbf{r}}
\newcommand{\Tx}{T_\chi}
\begin{document}

\vspace*{1.2cm}

\thispagestyle{empty}
\begin{center}

{\LARGE \bf Dark Matter in Stars}

\par\vspace*{7mm}\par

{

\bigskip

\large \bf Aaron C. Vincent}

\bigskip

{\large \bf  e-mail: aaron.vincent@queensu.ca}

\bigskip

{Department of Physics, Engineering Physics and Astronomy, \\ Queen's University, Kingston ON K7L 3N6, Canada} \\
{Arthur B. McDonald Canadian Astroparticle Physics Research Institute, \\Kingston ON K7L 3N6, Canada} \\
{Perimeter Institute for Theoretical Physics, Waterloo ON N2L 2Y5, Canada}

\bigskip

{\it Presented at the 3rd World Summit on Exploring the Dark Side of the Universe \\Guadeloupe Islands, March 9-13 2020}

\end{center}



\begin{abstract}
I review some key aspects of capture and possible observable effects of particle dark matter in stars. Focusing on the transport of heat from captured asymmetric dark matter, I outline existing computational methods, and the challenges that must be overcome to continue pushing the field forward.
\end{abstract}
  
\section{Introduction}
\label{S:intro}
Significant efforts are underway at underground laboratories around the world to detect the minute but telltale signatures of direct interactions between galactic dark matter (DM) and ordinary baryonic nuclei. Because --- by definition --- DM must be very weakly interacting, such searches must take place in well-shielded environments, where interference from cosmic rays, thermal noise and radiogenic backgrounds are as low as possible. Such direct detection (DD) experiments rely on elastic scattering between DM and target nuclei to provide a detectable signature in heat, ionization, scintillation, or a combination thereof. Weak couplings necessarily mean that DD experiments are limited by exposure, and with each subsequent generation, experiments have gotten larger. Currently, the strongest limits on spin-independent are set by XENON1T \cite{Aprile:2018dbl}, a 3500 kg liquid xenon detector, and planning has begun for hundred-ton scale argon and xenon experiments with the potential to pummel their way through the dreaded neutrino floor. 

  If present in the lab, elastic scattering between DM and nuclei must also occur in natural systems. The largest nearby target for such an effect is the Sun: at 2 $\times 10^{30}$ kg and exposure $t_\odot = 4.57$ Gyr, it constitutes a truly titanic (if noisy) detector. Indeed, if DM scattering off solar nuclei brings it below the local escape velocity, the DM will become gravitationally bound and settle into an equilibrium configuration near the core. Depending on the nature of the DM itself, it may then suffer one of three possible fates: 1) if it is too light, it will ``evaporate'' from momentum exchanges large enough to bring it above the local escape velocity\footnote{as long as it does not interact again on the way out \cite{Gould87a,Nauenberg87,Busoni:2017mhe}.} 2) if it is self-conjugate, or if sufficient quantities of ``anti-DM'' are present in the star, it will annihilate, or 3) if it is sufficiently heavy and asymmetric \cite{Zurek13}, it can act as a heat conductor \cite{Steigman78}, thanks to its long mean free path inside the solar plasma. 
  
  The latter two fates have observable consequences. DM annihilation into SM products and their subsequent decays into neutrinos can produce observable signals at underground (or under-ice) neutrino telescopes. Indeed, the strongest bounds on DM-nucleon scattering for certain DM candidates come from this channel. Heat transport can have more subtle consequences: by flattening the temperature gradient in the inner Sun, neutrino fluxes can be reduced, and the pressure and density profile of the Sun can be modified, changing helioseismology observables such as the convective zone radius $r_{CZ}$, the surface helium composition, and the inferred sound speed profile \cite{Lopes02a}. In other main sequence stars, convective cores can be erased, and with large enough concentrations, evolutionary trajectories on the Hertzsprung-Russel (HR) diagram can be severely modified.
  
In the following, I will focus on the latter effects, with special emphasis on some of the details of the calculations. However, I would be remiss not to mention that DM of various shapes and sizes can have even more spectacular consequences when combined with the exotic environments of white dwarfs or neutron stars, see e.g. \cite{PhysRevD.40.3221,Bertone:2007ae,Bramante:2015cua,Bramante:2017ulk,Acevedo:2019gre,Acevedo:2019agu} and references therein.

We shall start by recalling the capture rate of DM in stars, and briefly look at annihilation before turning our full attention to the perplexing problem of particle propagation and heat transport. 

\section{Capture and annihilation}
\label{sec:cap}
If the Milky Way's DM halo is near hydrostatic equilibrium, its velocity distribution in our vicinity should be roughly Maxwellian, with a dispersion velocity around 220 km/s, which can be obtained from the mass enclosed within the Sun's orbit. Though prior simulations cast doubt on this simple model, newer numerical simulations including the hydrodynamics of gas, star formation and feedback indicate that it is a fairly reasonable assumption \cite{Bozorgnia:2017brl}. The capture rate $C_\star$ of DM in a star of radius $R_\star$ is:
\begin{equation}
C_\star(t) = 4\pi \int_0^{R_\star} r^2 \int_0^\infty \frac{f_\star(u)}{u} w \Omega(w) \ud u\ud r.
\label{caprate}
\end{equation}
where $u$ is the DM speed in the star's frame, and $w(r) = \sqrt{u^2 + v_{esc}(r)^2}$, where $v_{esc}(r)$ is the escape velocity from a distance $r$ from the centre of the star. $f_\star(u)$ is the local DM speed distribution, and $\Omega(w)$ is a function that encodes the scattering kinematics: $w\Omega(w)$ it is proportional to the probability per unit time of a collision occurring that brings a DM particle with speed $w$ below $v_{esc}$. 

For the Sun, the only free parameters in \eqref{caprate} are the DM mass $m_\chi$ and the DM-nucleon\footnote{Interactions with electrons can also lead to capture, for a lower mass range \cite{Garani:2017jcj}.} cross section $\sigma \equiv d\sigma_{\chi-n}/dE_R$. The latter can result in non-trivial DM-\textit{nucleus} interactions. Depending on the Lorentz structure of the DM-quark vertex, the cross section depends generically on combinations of the non-relativistic quantum operators $\mathds{1}$ (the identity), $\vec q$ (the exchanged momentum), $\vec v^\perp$ (the relative velocity component orthogonal to $\vec q$), $\vec S_\chi$ and $\vec S_n$ (the DM and nucleon spins) \cite{Fitzpatrick13}. Each bilinear combination of these operators leads to both different kinematics and a different multipole projection onto the nuclear state, leading in turn to an isotope-dependent \textit{nuclear response}. These have been computed and tabulated in a number of references, including \cite{Catena:2015uha} in the context of the Sun. Operators that depend on $S_n$ are particularly interesting, as they do not benefit from the coherent enhancement $\sigma \propto A^2$ that spin-independent models do, and thus are much more difficult to probe with puny Earth-based detectors. Different scattering kinematic also mean that DD experiments probe very different areas of $q$ and $v$-space, leading to strong complementarity between approaches.

 Finally it is worth noting that $C_\star$ cannot be larger than the geometric limit set by the size of the stellar disk itself. This turns out to be larger than $\pi R_\star^2$ thanks to gravitational focusing. We point the interested reader to \textit{Capt'n General}\footnote{\url{https://github.com/aaronvincent/captngen}} \cite{Athron:2018hpc,NealinProg}, a set of numerical functions for calculating the capture of DM in stars including the above effects. We also note \cite{Lopes:2020dau} who explored the effects of general uncertainties in the DM velocity distribution.

If the DM then self-annihilates, annihilation products can produce high-energy neutrinos, detectable at Earth \cite{Silk:1985ax,1986PhLB..180..375H,Srednicki:1986vj,Gaisser:1986ha,Olive:1987av}. If the DM population attains an equilibrium between decay and annihilation, the neutrino production rate depends only on $\sigma$. For spin-independent interactions DD experiments are far more sensitive; however, for spin-dependent interactions, leading limits at high masses are set by SuperKamiokande \cite{Choi:2015ara} and IceCube \cite{Aartsen:2016zhm}. Before moving on to the main topic of heat transport, we point out the recent code $\chi$aro$\nu$ \cite{Liu:2020ckq} which self-consistently computes production and propagation of neutrinos from DM in the Sun.

\section{Heat transport: the Knudsen problem}
If the DM can accumulate in sufficient quantities, its small interaction cross section $\sigma \ll \sigma_T$ leads to measurable heat transport even for comparatively low DM populations (in the Sun, the local DM density means that  $m_\chi N_\chi \lesssim 10^{-10} M_\odot$). Computing the  observable effects of such heat transport requires implementation of the capture and transport calculations into a full Standard Stellar (Solar) Model (SSM) simulation such as GARSTEC \cite{weiss:2008} or MESA \cite{Paxton2011}, and evolving the star within the DM halo up to its current age $t_\odot$. SSM's typically have two free parameters: the initial helium density, and a mixing length parameter used to model convection in a 1d simulation. This means that the presence of an additional transport mechanism can still lead to solar models that satisfy the observed luminosity, age and radius. 

 As mentioned earlier, ADM can lead to a reduction (or spectral change \cite{Lopes:2018wgp}) of the $^8$B and $^{7}$Be neutrino fluxes from the Sun by an $O(1)$ fraction due to the lower central temperature without affecting the overall luminosity. Changes in structure also introduce effects on heliosesimological measures \cite{Lopes02a,Lopes:2012,Lopes:2014} including the radius of the convective zone boundary $r_{CZ}$, the sound speed profile $c_s(r)$ and dimensionless frequency separation ratios which can be constructed to probe the core composition without systematic effects from higher radii .  In slightly more massive stars than the Sun, the convective core can be erased by flattening the temperature profile so as to smoothly maintain local hydrostatic equilibrium across $r$. This has already be probed via aseteroseismological measurements \cite{Casanellas:2013,Casanellas:2015uga,Martins:2017fji}.

The computation of heat transport effects in stars is conceptually straightforward, but devilish in implementation. The phase space distribution $F(\uu,\rr,t)$ of captured DM follows a Boltzmann Collision Equation (BCE):
\begin{equation}
DF(\uu,\rr,t) = (\partial_t + \uu \cdot \nabla_\rr - \mathbf{g}(\rr) \cdot \nabla_\uu)F(\uu,\rr,t) = \frac{1}{l}CF(\uu,\rr,t),
\label{eq:BCE}
\end{equation}
where $\mathbf{g} = \nabla \phi$ is the local gravitational acceleration, $l$ is the typical interscattering distance and  $C$ is the collision operator. $CF(\uu,\rr,t)$ represents the scattering rate of DM with nuclei from any velocity to $\uu$ minus the scattering rate from $\uu$ to any other velocity. The microphysics of the DM-nucleus interactions are encoded in the collision operator --- see \cite{VincentScott2013} for a general treatment. Spherical symmetry and the fact that the equilibration time scale is much faster than the stellar evolution time scale (i.e. $\partial_t F \simeq 0$) simplify things a little bit. Alas, not nearly enough for comfort. 

Projecting the kinetic energy times the solution, $ (m u^2/2)F(\uu,rr)$, onto the radial direction, one arrives at the luminosity $L(r)$ carried through a shell at radius $r$ by DM. The energy deposited per unit stellar density $\rho$ per unit time is just:
\begin{equation}
\epsilon(r) = \frac{1}{4\pi r^2 \rho(r)}\frac{dL(r)}{dr}.
\label{eq:epsdef}
\end{equation}

Three approaches are generally available to us in tackling the BCE depending on the \textit{Knudsen number}  $K = l/r_\chi$, i.e. the ratio of the mean interscattering distance $l \sim 1/(\sigma n_{nuc})$ to the DM scale height in the star. 
\begin{figure}[ht!]
\includegraphics[width=\textwidth]{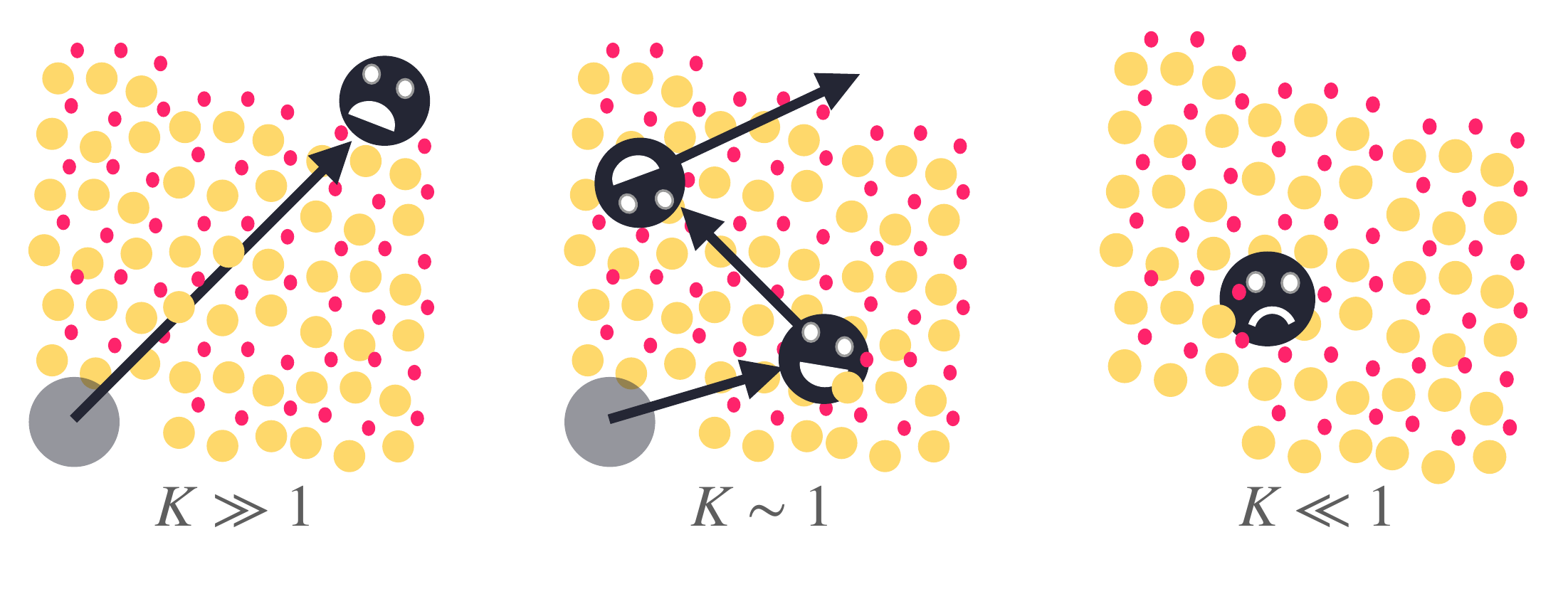}
\caption{Three heat conduction regimes by dark matter in the solar plasma. Left (Knudsen regime): large mean free paths (small $\sigma$) computable with the Spergel and Press (SP) approach lead to low overall energy deposition. Right (LTE regime): small mean free paths, computable with the Gould and Raffelt (GR) approach, mean the DM is ``stuck'' as $\sigma$ grows. Centre: at the Knudsen transition, heat transport is optimized. This regime does not have an analytical solution and must be calculated with a Monte Carlo-calibrated interpolation. The SP solution is based on incorrect assumptions, and GR can be numerically unstable and breaks down at small radii.}
\end{figure}
\begin{enumerate}
\item In the weakly-interacting regime, $CF$ is very small, giving a simple solution to the BC(ollisionless)E $DF = 0$: 
\begin{equation}
n_{\chi,\mathrm{iso}} \propto e^{-\phi(r)/T_\chi},
\label{eq:nxiso}
\end{equation}
where the DM temperature $T_\chi$ is a weighted average of the temperatures of the heat bath the DM interacts with. After ``some algebra'' , the transported energy \eqref{eq:epsdef} was obtained by \textbf{Spergel and Press} (SP \cite{Spergel85}):\footnote{I have omitted the requisite sum over nuclear species to keep this equation on a single line. Pretend that it is there.}
\begin{equation}
\begin{array}{c}{\epsilon_{\mathrm{SP}}\left(r, t, \Tx\right)=\frac{8 \sqrt{\frac{2}{\pi}} \mathrm{k}^{3 / 2}}{\rho_{\star}(r, t)} n_{\chi, \mathrm{iso}}(r, t)\left[T_{\star}(r, t)-\Tx(t)\right]}  {  \sigma n_{nuc}(r, t) \frac{m_{\chi} m_{\mathrm{nuc}, i}}{\left(m_{\chi}+m_{\mathrm{nuc}}\right)^{2}}\left(\frac{T_{\star}(r, t)}{m_{\mathrm{nuc}}}+\frac{\Tx(t)}{m_{\chi}}\right)^{1 / 2}}\end{array}.
\end{equation}
Note that $\epsilon$ gets weaker with smaller $\sigma$, as the interaction rate becomes smaller. This looks very thermodynamicsy, but the inconsistent assumption that $CF = 0$ will turn out to be \cite{Nauenberg87} one of the reasons that this treatment will yield inaccurate results. 
\item In the \textit{Local Thermal Equilibrium} $K \ll 1$ regime, the DM is locally at the same temperature at the nuclei. This allowed \textbf{Gould and Raffelt} (GR,  \cite{GouldRaffelt90a})\footnote{This builds on earlier work Faulkner \& Gilliland \cite{Faulkner85} and Gilliland et al. \cite{Gilliland86}. } to expand the BCE to first order in the small quantity $\varepsilon = l(r) |\nabla \log T(r)|$:
\begin{equation}
F(v,r) = F_0 + \varepsilon \cdot dipole,
\end{equation}
where $F_0$ is again the Maxwell-Boltzmann solution to $DF = 0$ but with $T(r)$ equal to the local stellar temperature, and the \textit{dipole} contribution is responsible for the local flux of heat due to DM. This allows for the computation of two quantities that depend only on $\mu = m_\chi/m_{nuc}$, the ratio of the DM to nucleon masses. These are a molecular diffusion coefficient $\alpha(\mu)$ (or ``fluffiness parameter'') that governs the DM radial distribution, and $\kappa(\mu)$\footnote{$\kappa$ is a function of $r$ in Eq. \eqref{eq:Llte}. This is because the isotopic abundances, which govern the average value of $\mu$, are radially-dependent.}, a thermal conductivity coefficient. The luminosity is
\begin{equation}
L_{LTE} = 4 \pi r^2 n_\chi(\alpha,r) l(r) \kappa(r) \sqrt{\frac{T}{m_\chi}}\frac{dT}{dr},
\label{eq:Llte}
\end{equation}
And $\epsilon$ is obtained via Eq. \eqref{eq:epsdef}.

 In contrast with the SP solution, the LTE solution becomes weaker with increasing cross section.
  
\item A direct Monte Carlo simulation can yield an equilibrium solution of the BCE, as the set of phase space coordinates sampled in the long time limit in a static background plasma is ergotically equivalent to a large collection of particles in equilibrium. While this allows for an exact solution of the BCE in principle, it is practically infeasible as it requires a separate simulation for every set of DM parameters, and for every evolutionary time step in the star's lifetime. Still, it may be used to validate the above approaches: this was done by Gould \& Raffelt \cite{GouldRaffelt90a,GouldRaffelt90b}, who notably concluded that the isothermality assumption in the SP approach indeed leads to an incorrect luminosity curve, and while the GR calculation yields accurate results over most of the star in the LTE regime, the luminosity at low radii is overestimated in both cases, because the isotropy assumption in $\vv$ breaks down near $r = 0$. 
\end{enumerate}
The ``correct'' technique that is accepted and widely used today is the GR (LTE) technique, rescaled with a ``Knudsen correction'' based on the GR MC simulations that recovers the correct behaviour in the large $K$ regime, and a ``radial correction'' that accounts for the isotropy effects \cite{Bottino02} by suppressing luminosity at low $r$. 

\begin{figure}
\begin{center}
\begin{tabular}{c c}
\includegraphics[width=0.5\textwidth]{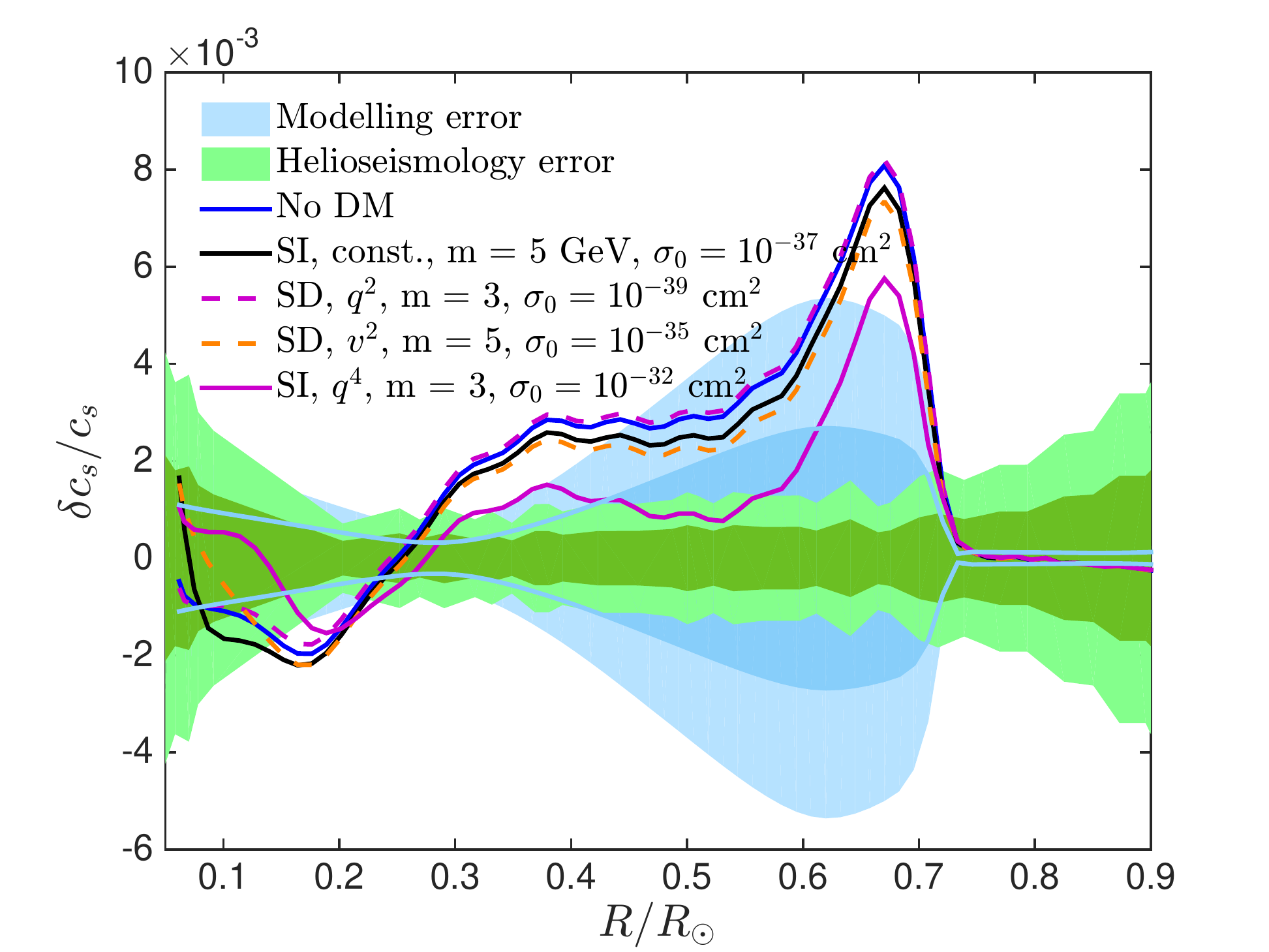}&\includegraphics[width=0.53\textwidth,trim=0 .8cm 0 0cm]{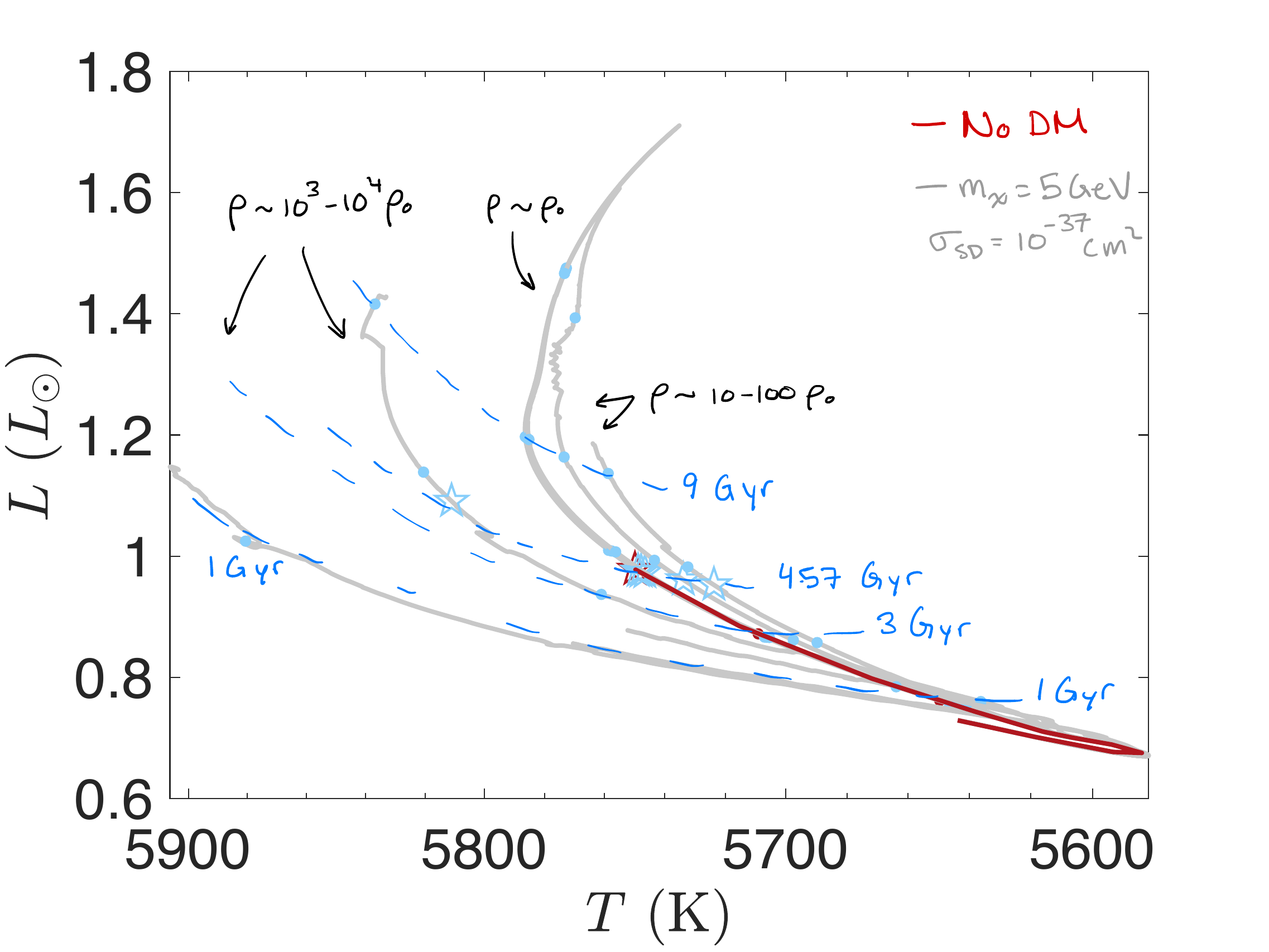}
\end{tabular}
\end{center}
\caption{Left: Figure from \cite{Vincent:2016dcp}. Improvement of the difference in the radial sound speed profile $c_s(r)$ between standard solar models without (blue) and with the capture and heat transport of asymmetric dark matter (other colored lines). Bands represent 1 and 2 $\sigma$ modelling (blue) and helioseismology (green) errors.  Right: Luminosity-Temperature plot showing the evolution of a 1 $M_\odot$ star capturing ADM with different local densities (grey bands). Approximate lines of constant age are shown in dashed blue. }
\label{fig:cs}
\end{figure}

The effect of ADM heat transport is largest for DM masses that are best kinematically-matched with H and He, while heavy enough to avoid evaporation: $m \sim 3-5$ GeV. The most interesting effects unsurprisingly occur near the Knudsen transition. For a constant DM-nucleon cross section this is around $10^{-35}$ \cmtwo for spin-dependent interactions, and  $10^{-37}$ \cmtwo in the spin-independent case. While these fall above upper limits set by earth-based DD experiments, non-constant interactions $\sigma \propto v^n, q^n$ ($n = -2, 2, 4$) \cite{VincentScott2013,Vincent:2014jia,Vincent:2015gqa,Vincent:2016dcp} as well as some theoretically-motivated models \cite{Lopes14,Geytenbeek:2016nfg} can give Knudsen transitions for values of the cross section that are compatible with DD limits. Such models can also improve on the SSM by up to 6$\sigma$, providing a possible path \cite{Frandsen:2010yj,Taoso10,Cumberbatch10} to resolving the \textit{Solar Composition Problem}, a strong disagreement between helioseismological observables and SSMs \cite{2005ASPC..336...25A,Asplund:2009fu,Serenelli:2009yc}. The left panel of Fig \ref{fig:cs} shows the improvement in the sound speed profile for a variety of dark matter models with $\sigma = \sigma_0 (v/v_0)^n$ or  $\sigma = \sigma_0 (q/q_0)^n$. These can be related with the NREO models described earlier.

Thanks to DD experiments, the parameter space is rapidly closing, and we may well be forced deep into the Knudsen regime. Here, solar effects may be more difficult to observe, but stars near the galactic centre (GC) that can capture far more DM over their lifetimes can still serve as competitive probes of new physics for small values of $\sigma$. Indeed, large amounts of DM can affect the relation between a star's mass, luminosity and temperature by changing the conditions of local thermal and hydrostatic equilibrium. This ultimately means that a star's trajectory on the main sequence can be very different from the standard prediction \cite{Scott09,Zentner:2011wx,Iocco12,Hurst:2016rfj,2019ApJ...879...50L}. This is where trouble arises. There are three issues at play:
\begin{enumerate}
\item The GR formalism relies on two numerical derivatives, $ L \propto  dT/dr$ and $\epsilon \propto dL/dr$. Modern stellar evolution codes typically contain small discontinuities in their temperature profiles which are ordinarily not a problem. However, when DM heat transport is large, these discontinuities are amplified and can yield wild, unphysical self-amplifying oscillations for interstellar DM densities larger than:
\begin{equation}
\log\left(\frac{\rho}{\rm GeV}\right) \gtrsim 0.5 - 2\log\left(\frac{\sigma}{10^{-37} \mathrm{cm^2}} \right).
\end{equation}
We refer to this region as the \textit{Danger Zone} \cite{LukeinProg,Loggins}.
\item As we are deep in the Knudsen regime, it becomes increasingly unsettling to use an extrapolation of the GR formalism which \textit{was developed using the explicit assumption of small mean free paths}.
\item The alternative approach, SP, is not self-consistent and does not agree with Monte Carlo simulations.
\end{enumerate}
In order to progress beyond point 1., many references have nonetheless gone ahead and obtained interesting results using the SP approach. These lead to interesting and suggestive results: increased heat transport leads to changes in the HR evolution of MS stars, notably erasing convective cores and significantly extending their main sequence lifetime \cite{Zentner:2011wx,Iocco12,Hurst:2016rfj,2019ApJ...879...50L}. On a color-magnitude diagram, this means a modification of the MS turnoff that depends on the local density of DM. I show this in the right panel of Fig. \ref{fig:cs}, which shows a number of trajectories for a 1 $M_\odot$ star interacting with DM with densities varying from 1-10$^4$ times the local DM density $\rho_0 = 0.4$ GeV cm$^{-3}$, produced using the MESA \cite{Paxton2011} stellar evolution software and the SP approach.

These conclusions are likely to be fairly robust, even if they are built on shaky theoretical foundations. But the above objections should emphasize the fact that more work is needed if we are to use stars and stellar populations not only as a probe for the effects of dark matter, but as a way to measure the DM properties themselves. The way forward is twofold: 1) revisiting the BCE from the non-local point of view; and 2) careful comparison with state-of-the-art Monte Carlo simulations.


\section{Conclusions}
The night sky is strewn with thousands upon thousands of free, ultra-massive dark matter direct detection experiments. As more precise observations make it possible to perform asteroseismology on individual stars and quality population analyses, better computational techniques will be needed to  accurately predict the impact of DM on these stars in the hopes of advancing in our quest for knowledge of the dark side.

\section*{Acknowledgements}
I thank Air Canada and the Canadian Government for waiting until the day after I returned home before halting all travel in March of 2020. Special thanks to Pierre Petroff, Betty Calpas and the other members of the local organizing committee for EDSU2020. I acknowledge support from the Arthur B. McDonald Institute, CFI (Canada) and MEDJCT (Ontario). 

\bibliographystyle{JHEP_pat} 
\bibliography{references.bib}


\end{document}